\documentclass[conference,10pt]{IEEEtran}
\IEEEoverridecommandlockouts
\usepackage{cite}
\usepackage{microtype}
\ifCLASSINFOpdf
  \usepackage[pdftex]{graphicx}
  \DeclareGraphicsExtensions{.pdf,.jpeg,.png,.JPG}
\else
  \usepackage[dvips]{graphicx}
  \DeclareGraphicsExtensions{.eps}
\fi
\graphicspath{{./graphics/}}
\usepackage{mathtools}
\usepackage[caption=false,font=footnotesize]{subfig}
\usepackage{url}
\usepackage{booktabs}
\usepackage{siunitx}
\usepackage{xcolor}

\newcommand{\FAU}{Friedrich-Alexander-Universit\"at Erlangen-N\"urnberg}
\newcommand{\SEON}{ENT-Department, Section of Experimental Oncology and Nanomedicine (SEON)}
\newcommand{\UK}{Universit\"atsklinikum Erlangen}
\newcommand{\LTE}{Institute for Electronics Engineering}
\newcommand{\IDC}{Institute for Digital Communications}
\newcommand{\EFI}{Emerging Fields Initiative}
\newcommand{\vmax}{v_{0}}
\newcommand{\veff}{v_\mathrm{eff}}
\newcommand{\Pob}{P_\mathrm{ob}}

\newcommand{\tpeak}{t_\mathrm{peak}}
\newcommand{\Pec}{\mathrm{Pe}}
\newcommand{\Reyn}{\mathrm{Re}}

\newcommand{\Qb}{Q_\mathrm{b}}
\newcommand{\Qp}{Q_\mathrm{p}}
\newcommand{\Vi}{V_\mathrm{i}}
\newcommand{\chiref}{\chi_\mathrm{ref}}
\newcommand{\arx}{a_\textsc{rx}}
\newcommand{\Vrx}{V_\textsc{rx}}

\begin{document}
\title{Experimental Molecular Communication Testbed Based on Magnetic Nanoparticles in Duct Flow}

\author{\IEEEauthorblockN{Harald Unterweger\IEEEauthorrefmark{1},
Jens Kirchner\IEEEauthorrefmark{2},
Wayan Wicke\IEEEauthorrefmark{3}, 
Arman Ahmadzadeh\IEEEauthorrefmark{3},
Doaa Ahmed\IEEEauthorrefmark{2},
Vahid Jamali\IEEEauthorrefmark{3}, \\
Christoph Alexiou\IEEEauthorrefmark{1},
Georg Fischer\IEEEauthorrefmark{2},
and Robert Schober\IEEEauthorrefmark{3}}
\IEEEauthorblockA{\IEEEauthorrefmark{1}\SEON, 
    \UK
}
\IEEEauthorblockA{\IEEEauthorrefmark{2}\LTE, \FAU
}
\IEEEauthorblockA{\IEEEauthorrefmark{3}\IDC, \FAU}
\thanks{%
    This work was supported in part by the \EFI{} (EFI) of the \FAU{} (FAU).%
}%
}

\IEEEspecialpapernotice{(Invited Paper)}

\maketitle

\begin{abstract}
Simple and easy to implement testbeds are needed to further advance molecular communication research.
To this end, this paper presents an in-vessel molecular communication testbed using magnetic nanoparticles dispersed in an aqueous suspension as they are also used for drug targeting in biotechnology.
The transmitter is realized by an electronic pump for injection via a Y-connector.
A second pump provides a background flow for signal propagation.
For signal reception, we employ a susceptometer, an electronic device including a coil, where the magnetic particles move through and generate an electrical signal.
We present experimental results for the transmission of a binary sequence and the system response following a single injection.
For this flow-driven particle transport, we propose a simple parameterized mathematical model for evaluating the system response.
\end{abstract}

\IEEEpeerreviewmaketitle

\section{Introduction}
Applications such as drug targeting or monitoring of chemical reactors spurred the interest in and theoretical growth of molecular communication; an approach for communication using small particles in areas impenetrable for electromagnetic waves, see \cite{Farsad_comprehensive_2016} for an overview of the recent literature.

Experimental studies have successfully demonstrated different components of a molecular communication system, see \cite{Krishnaswamy_Time_2013,Nakano_Microplatform_2008,Nakano_Performance_2017,Felicetti_Modeling_2014,Abbasi_Controlled_2018} and references therein.
However, realizing a fully-functional artificial molecular communication system at nanoscale remains a challenge.
Nevertheless, for first experimental insights, testbeds in the size range of several \si{\centi\meter} have been proposed.
In particular, the system described in \cite{Farsad_Tabletop_2013} is based on spraying and detecting alcohol in open space, and the testbed in \cite{Farsad_Novel_2017} is based on signaling with acids and bases for signaling within closed vessels.
The testbed in \cite{Farsad_Tabletop_2013} has been extended to a multiple-input and multiple-output (MIMO) system and to a confined environment within a metal pipe \cite{Koo_Molecular_2016,Qiu_molecular_2014}.
Also, improved theoretical models have been proposed to account for apparent discrepancies between theory and experiment \cite{Farsad_Channel_2014}.

Considering the potential applications, a testbed using tubes, e.g., for emulating blood vessels, is relevant.
On the other hand, using chemicals like acids and bases for information transmission could potentially interfere with other processes in an application environment, e.g., in the body.
Furthermore, the detection mechanism in \cite{Farsad_Novel_2017} is intrusive as it relies on a pH-electrode inserted into the vessel.
On the other hand, information particles do not have to be restricted to those occurring in nature \cite{Farsad_comprehensive_2016}.
One type of artificial particle that is already well-established in biotechnology are biocompatible magnetic nanoparticles \cite{Pankhurst_Applications_2003}.
These particles can be tailored to a particular application by engineering of their size, composition, and coating \cite{Lu_Magnetic_2007}.
Moreover, magnetic nanoparticles can be attracted by a magnet and externally visualized, which can help detection and supervision.
Applications of magnetic nanoparticles include tissue engineering, biosensing, imaging, remotely stimulating cells, waste-water treatment \cite{Raj_Coconut_2015}, and drug delivery, see \cite{Pankhurst_Applications_2003} and references therein.

In the context of molecular communication, the use of magnetic nanoparticles has been considered in \cite{Wicke_Molecular_2017} and \cite{Kisseleff_Magnetic_2017}, where the benefits of attracting them as information carriers towards a receiver are evaluated and a wearable device for detecting the presence of magnetic nanoparticles is proposed, respectively.
However, a practical molecular communication testbed employing magnetic nanoparticles has not been reported, yet.

Similar to \cite{Farsad_Novel_2017}, in this paper, we present a testbed for in-vessel molecular communication.
Our setup differs in that it uses specifically designed magnetic nanoparticles as information carriers, which are biocompatible, clinically safe and do not interfere with other chemical processes like acids and bases would do, and thus might be attractive for applications such as the monitoring of chemical reactors where particles stored in a reservoir could be released upon an event like the detection of a defect.
Here, magnetic nanoparticles are injected and transported along a propagation tube using two electronic pumps.
The propagation tube leads through the receiver where the magnetic susceptibility of the mixture of water and magnetic nanoparticles within a tube section can be non-intrusively determined.
The magnetic susceptibility measured at a tube section is proportional to the concentration of the particles within the section.
This proportionality lends itself better for mathematical analysis than the pH, which depends on the underlying proton concentration in a more complicated manner~\cite{Farsad_Novel_2017}.

For the chosen parameters, we find laminar flow-driven particle transport applicable for signal propagation after injection.
We model the injection as being axially concentrated and adhering to a parameterized initial distribution in the cross section at the site of injection.
Modeling this transport by neglecting diffusion and assuming a parabolic flow profile yields a good agreement with the experimental results.

The rest of this paper is organized as follows.
In Section~\ref{sec:system_description}, we describe the components of the proposed system and the overall testbed.
In Section~\ref{sec:system_analysis}, we characterize the expected system response mathematically.
In Section~\ref{sec:experiments}, we present experimental data and evaluate the theoretical model.
Finally, Section~\ref{sec:conclusion} concludes the paper.

\section{Magnetic Nanoparticle Based Testbed}
\label{sec:system_description}
\begin{figure*}[!t]
    \centering
    \subfloat[]{
        \includegraphics{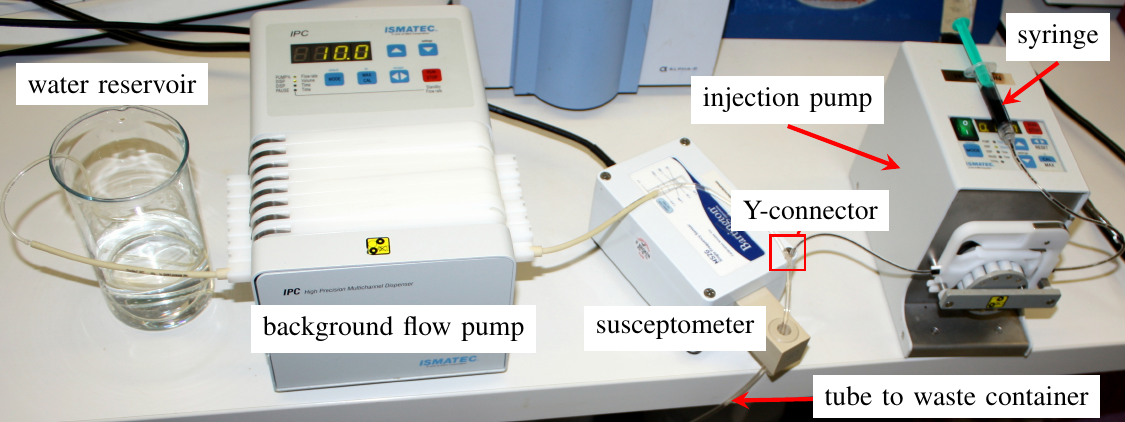}
        \label{fig:aufbau}
    }
    \subfloat[]{
        \includegraphics[]{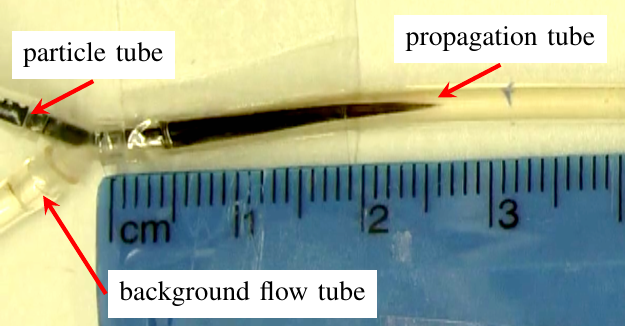}
        \label{fig:injection}
    }
    \caption{
        (a) Photograph of the testbed showing the water reservoir, the background flow pump, the susceptometer, the pump used for injection, the syringe holding the suspension of SPIONs, and flexible plastic tubes connecting the components.
        The waste container below the table is not shown.
        (b) Photograph of the Y-connector with elongated particle suspension right after injection for a slow background flow of $\Qb=\SI{1}{\milli\liter/\minute}$.
        Ruler with \si{\centi\meter} scale for reference.
    }
    \vspace*{-0.25cm}
\end{figure*}

In the following, each component of the system is briefly described.
A representative photograph of the whole system is shown in Fig.~\ref{fig:aufbau}, while Table~\ref{tab:system parameters} summarizes the system parameters.

\begin{table}
    \caption{System Parameters}
		\label{tab:system parameters}
    \centering
    \begin{tabular}{lr}
        \toprule
        Parameter                            & Numerical Value\\
        \midrule
        Hydrodynamic particle radius                   & \SI{27.5}{\nano\meter}\\
		Suspension iron stock concentration            & \SI{7.89}{\milli\gram/\milli\liter}\\
        Suspension magnetic susceptibility             & \SI{7.28d-3}{} (SI units) \\
        \midrule
        Tube radius particle injection      & \SI{0.40}{\milli\meter}	\\
        Tube radius background flow	$a$	    & \SI{0.75}{\milli\meter}	\\
        Flow rate particle injection $\Qp$  & \SI{5.26}{\milli\liter/\minute}	\\
        Flow rate background flow $\Qb$	    & \SI{5}{\milli\liter/\minute}	\\
        Volume particle injection $\Vi$		& \SI{14}{\micro\liter}	\\
        Duration particle injection		    & \SI{160}{\milli\second} \\
        Binary symbol duration $T$		    & \SI{4}{\second}	\\
        Propagation distance $d$            & \SI{5}{\centi\meter}\\
        Receiver length $c_z$			    & \SI{18}{\milli\meter} \\
        Receiver inner radius $\arx$	    & \SI{5}{\milli\meter} \\
        \bottomrule
    \end{tabular}
    \vspace*{-0.25cm}
\end{table}

\subsection{Carrier and Transmitter}
In the considered system, superparamagnetic iron oxide nanoparticles (SPIONs), which were originally developed for biomedical applications, are used as information carriers, due to their magnetic properties.
In particular, we employed lauric acid coated SPIONs (SPION\textsuperscript{LA}), which were originally developed for magnetic drug targeting purposes \cite{Zaloga_lauric-acid_2014}. 
The particles are dispersed in an aqueous suspension and stored in a syringe, which is connected to a tube with an inner radius of \SI{0.4}{\milli\meter}.
These particles have a hydrodynamic radius of \SI{27.5}{\nano\meter}, an iron stock concentration of \SI{7.89}{\milli\gram/\milli\liter}, a susceptibility of \num{7.28d-3} (dimensionless in SI units), and a concentration of
approximately \SI{4d13}{Particles/\milli\liter} in aqueous suspension.
The movement of the particles through the tube is established with a computer controlled peristaltic pump (Ismatec\textsuperscript{\textregistered} Reglo Digital, Germany), which can provide discrete pumping actions at a flow rate of \SI{5.26}{\milli\liter/\minute}, injecting a dosage volume of \SI{14}{\micro\liter} of particle suspension.

The end of the tube with the particles is joined via a Y-connector with another tube of radius \SI{0.75}{\milli\meter} providing a background flow, see Fig.~\ref{fig:injection}.
The constant background flow of water has a flow rate of \SI{5}{\milli\liter/\minute} and is maintained by a second pump (Ismatec\textsuperscript{\textregistered} IPC, Germany).

\subsection{Propagation Channel}
\label{sec:system_description_channel}
The Y-connector constitutes the end of the transmitter; its end piece passes into the propagation channel, which also has an inner tube radius of \SI{0.75}{\milli\meter}.
The flow rate in this channel is the sum of the rates of the background flow and the particle injection.
It is hence time-dependent and amounts to \SI{10.26}{\milli\liter/\minute} during particle injection and \SI{5}{\milli\liter/\minute} in the remaining time.
When particles are pumped into the channel by the transmitter, then the resulting particle cloud is entrained by the flow and simultaneously diluted and elongated, see Fig.~\ref{fig:injection}.

The length of the propagation channel is variable, but was set to \SI{5}{\centi\meter} for the results shown.

\subsection{Receiver}
At the end of the propagation channel, the tube runs through the air core of an MS2G Bartington\textsuperscript{\textregistered} susceptometer coil (inner diameter: \SI{10}{\milli\meter}, height: \SI{5}{\milli\meter}).
When the magnetic particles are within the detection range of the susceptometer, an electrical signal $\chi(t)$ is generated.
This signal is proportional to the number of SPIONs that are within the detection range at a specific time instance.
After the particles have passed through the receiver, they are collected in a waste bin together with the water from the background flow.
Water has a small negative magnetic susceptibility of about \SI{-9.04e-6}{} (SI units) \cite{White_Fluid_2010}.
However, its magnitude is much less than that of the considered particle suspension $\chiref=\SI{7.28e-3}{}$ (SI units).

\subsection{Communication Scheme}
\label{sec:system_description_communication}
Modulation uses on-off keying and is realized by a custom LabVIEW (National Instruments, Austin, Texas, USA) graphical user interface (GUI) that triggers the discrete pumping actions of the particle pump: Every time a binary symbol ''$1$'' shall be sent by the transmitter, a dose of particle solution is injected into the propagation channel; if a ''$0$'' shall be sent, no particles are injected.
From the injection volume and the flow rate in the particle tube, the injection duration is calculated as \SI{160}{\milli\second}, while the symbol interval was set to \SI{4}{\second}.

For sending a text message, the 8 bit extended ASCII encoding for capital letters is used.
The 26 capital letters each have a $[0,1,0]$ prefix, which we use for synchronization.
In this way, the receiver recognizes the start of a character by the first detected peak position.

The susceptibility changes measured at the receiver were recorded by use of the software Bartsoft 4.2.1.1 (Bartington Instruments, Witney, UK) provided by the manufacturer of the susceptometer.
For detection of SPION injections at the receiver, a constant threshold was applied.

Transmission proceeds as follows.
For each 8-bit message, the initial peak position $t_0$ is determined.
Then, the following five bits are detected by comparing $\chi(t_0+T+kT)$, $k=1,2,3,4,5$ with a threshold.

\section{System Model}
\label{sec:system_analysis}
In this section, we develop a simple model for the particle transport in the described testbed.
First, we determine characteristic dimensionless parameters relevant for our analysis.
Then, we briefly describe the particle transport and give an analytical expression for the expected system response.

\subsection{General Considerations}
Fluid flow can be categorized as either laminar or turbulent.
This categorization determines the appropriate mathematical model to be used.
While laminar flow is prevalent in microfluidic applications, turbulent flows are encountered in macroscale ducts in the size range of several \si{\centi\meter}.
The relevant parameter, in this case, is the Reynolds number $\Reyn$ which predicts a transition from laminar to turbulent flow within a circular duct at a value around \num{2100}, see \cite[Chapter~4.10]{White_Fluid_2010}.
Thereby, $\Reyn$ can be defined as \cite[Chapter~4.10]{White_Fluid_2010}
\begin{equation}
    \Reyn = \frac{2a \cdot \veff}{\nu},
\end{equation}
where $2a$ is the tube diameter, $\veff=\Qb/(\pi a^2)$ is the area-averaged velocity in the tube, and $\nu$ is the kinematic viscosity of the liquid in \si{\meter^2\per\second}.
For the parameters in Table~\ref{tab:system parameters}, we find $\veff=\SI{47.2}{\milli\meter/\second}$.
Thus, using the kinematic viscosity of water $\nu=\SI{e-6}{\meter^2/\second}$ \cite{White_Fluid_2010}, we obtain $\Reyn=\SI{70.7}{}<\num{2100}$ and hence expect fully laminar flow.

For laminar flow in a straight tube of circular cross section, the non-uniform flow velocity profile is well known to be \cite{White_Fluid_2010}
\begin{equation}
    \label{eq:flow profile}
    v(r) = \vmax \cdot \left( 1 - \frac{r^2}{a^2} \right),
\end{equation}
where $\vmax$ is the velocity in the center of the tube.
For this velocity profile, the area-averaged velocity $\veff$ can be obtained as $\veff=\vmax/2$.
In our testbed, the tube is not fully straight.
Nevertheless, as the deviations over regular distances on the order of the inner tube diameter are small, the flow profile \eqref{eq:flow profile} can be assumed to be a valid approximation.

In general, the particle transport is governed by both diffusion and the fluid flow described in \eqref{eq:flow profile}.
The relative importance of diffusion over the transport by fluid flow considering a distance of $d$ can be quantified by the P\'eclet number $\Pec$.
When $\Pec\gg d/a$ and $\Pec\ll d/a$, flow and diffusion dominate the transport, respectively.
Thereby, $\Pec$ can be defined as~\cite[Eq.~(4.6.8)]{Probstein_Physicochemical_2005}
\begin{equation}
    \Pec = \frac{a\cdot \veff}{D},
\end{equation}
where $D$ is the diffusion coefficient of the magnetic particles and can be estimated to be less than \SI{e-11}{\meter^2/\second} for the considered SPIONs.
Hence, for the parameters in Table~\ref{tab:system parameters}, we obtain $\Pec>\num{3.54e6}$.
This value is several orders of magnitude larger than $d/a=\num{66.7}$ and therefore the impact of diffusion is assumed to be negligible over the considered distance $d$~\cite{Probstein_Physicochemical_2005}.

Motivated by this dimensional analysis, for simplicity, we will assume the transport can be described only by the velocity profile in \eqref{eq:flow profile}, see also the shape of the propagating SPION cloud in Fig.~\ref{fig:injection}.

\subsection{Mathematical Model}

The flow at the injection site is complicated as highlighted in Section~\ref{sec:system_description_channel}.
Nevertheless, right after the injection pump stops pumping, the resulting volume distribution completely determines the received signal via the laminar flow in \eqref{eq:flow profile}.
For example, when due to the injection more particles are concentrated along the axis than on the boundary of the tube, a faster decay of the received signal and a larger peak can be expected.
In fact, in our experiments, we observed varying decay profiles of the received signal depending on the choice of parameters.
Hence, we are interested in developing a model for the initial volume distribution.
To this end, we note that regarding the received signal, the injected volume distribution is not unique.
In fact, due to the radially symmetric flow profile and the receiver covering the whole cross section of the tube, for any non-symmetric volume distribution there is always an equivalent radially symmetric volume distribution resulting in the same received signal.
Moreover, as first order approximation, we will assume that the initial distribution can be modeled as being axially concentrated at the site of injection as the time of injection is short compared to the symbol duration.
Then, for choosing this initial distribution, we have the following requirements.
First, it should have a parameter which can be tuned to the application scenario.
Second, it should lead to a simple model for the received signal while providing a good fit to the experimentally observed data.
We note that a more accurate model could be obtained, for example, by numerical simulation and evaluating the obtained initial volume distribution.
As this numerical simulation does not directly give theoretical insight, in this paper, we will focus on the transport by the laminar flow \eqref{eq:flow profile} and leave a more careful study of the injection process for future work.

Two common models for introducing particles in a straight tube are uniformly over the inlet cross section or in a concentration proportional to the velocity profile in \eqref{eq:flow profile} \cite[Chapter~15]{Levenspiel_Chemical_1999}.
This leads to observed particle concentrations at the outlet with decays proportional to powers of $1/t$ \cite[Chapter~15]{Levenspiel_Chemical_1999}.
Motivated by our experimental observation of varying decays of the received signal and the considerations above, we consider the following example initial SPION distribution over the cross section at the axial position of the injection
\begin{equation}
    \label{eq:initial distribution}
    f_{x,y}(x,y) = \frac{\beta+1}{\pi a^2} \cdot \left(1 - \frac{x^2+y^2}{a^2}\right)^\beta,
\end{equation}
where $\beta\geq0$ is a shape parameter and the Cartesian coordinates $x,y$ satisfy $x^2+y^2\leq a^2$.
For $\beta=0$, we obtain a uniform initial distribution $f_{x,y}(x,y) = 1/(\pi a^2)$ over the cross section.
For $\beta\to\infty$, we obtain $f_{x,y}(x,y)=\delta(x)\delta(y)$ where all particles are concentrated in the center of the tube.
In general, the larger $\beta$, the more particles are initially centered in the tube.

By the methodology in \cite{Wicke_Modeling_2017}, with \eqref{eq:initial distribution} we obtain the system response as
\begin{equation}
    \label{eq:Pob}
    \Pob(t) =
    \begin{dcases}
        0, & t\leq\frac{d}{\vmax} \\
        1 - \left(\frac{d}{\vmax t}\right)^{\beta+1}, & \frac{d}{\vmax} < t < \frac{d+c_z}{\vmax} \\
        \frac{(d+c_z)^{\beta+1} - d^{\beta+1}}{(\vmax t)^{\beta+1}}, & t\geq\frac{d+c_z}{\vmax}
    \end{dcases}
\end{equation}
which has a peak of height $1-(1+c_z/d)^{-1-\beta}$ for $\tpeak=(d+c_z)/\vmax$.
From the fit of \eqref{eq:Pob} to experimental results, we can infer the value of $\beta$ and the initial release distribution.
From \eqref{eq:Pob}, we obtain the expected susceptibility over time as
\begin{equation}
    \label{eq:system response}
    \chi(t) = \chiref \frac{\Vi}{\Vrx} \Pob(t),
\end{equation}
where $\chiref$ and $\Vi$ can be found in Table~\ref{tab:system parameters} and $\Vrx$ is the volume of the susceptometer's sensitive region which can be obtained as $\Vrx=c_z \pi\arx^2$, where $\arx$ and $c_z$ are the radius and the length of the receiver coil, respectively.

\section{Experimental Characterization}
\label{sec:experiments}
In this section, we present some experimental results obtained with the testbed and numerically evaluate the analytical model proposed in Section~\ref{sec:system_analysis}.
In the following, the parameters in Section~\ref{sec:system_description} apply unless specifically indicated otherwise.

\begin{figure}[!t]
    \centering
    \includegraphics[]{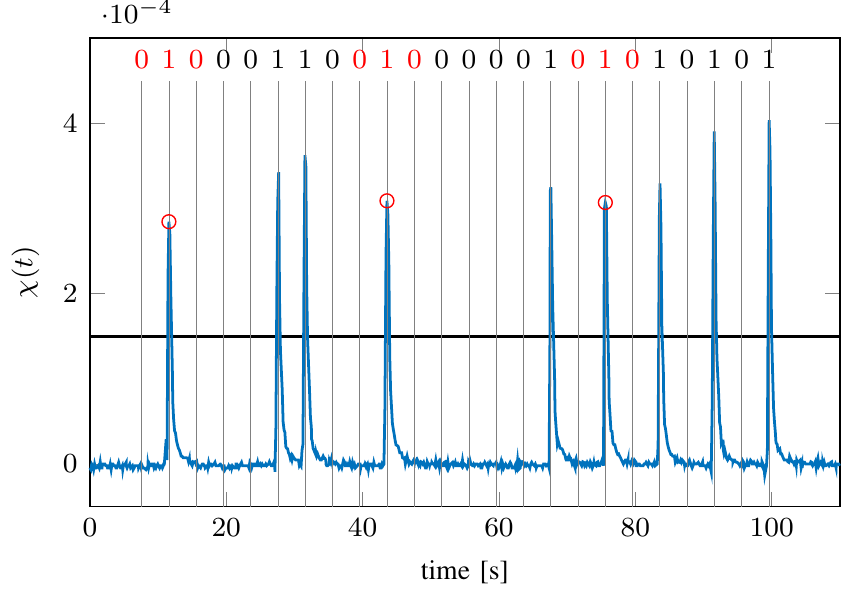}
    \caption{Example received signal for the bit sequence $[0,1,0,0,0,1,1,0,0,1,0,0,0,0,0,1,0,1,0,1,0,1,0,1]$.}
    \label{fig:example_sequence}
    \vspace*{-0.25cm}
\end{figure}

In Fig.~\ref{fig:example_sequence}, we show the measured received susceptibility time signal for the example bit sequence $[0,1,0,0,0,1,1,0,0,1,0,0,0,0,0,1,0,1,0,1,0,1,0,1]$ encoding 3 characters as described in Section~\ref{sec:system_description_communication}.
We also show the decoded bit sequence and the corresponding sampling times.
The threshold used for detection is set to $\chi_0=\SI{1.75e-4}{}$ and shown as horizontal black line.
Furthermore, the peaks used for synchronization, which occur at regular distances of $8T$ after the first peak, are marked with red circles.
Thereby, the very first peak around \SI{17}{\second} was found numerically by analyzing the whole received signal after the transmission of all bits.
In particular, all subsequent sampling times are shown by vertical subdued black lines.
The shown decoded bits evaluate to $1$ and $0$ if the received signal at the sampling times is above or below the threshold, respectively.
Comparing with the transmitted sequence, we observe that the transmitted sequence is perfectly recovered.
In fact, the received signal follows a straightforward pattern, where no pulses are observed when a $0$ was transmitted.
In other words, there is negligible intersymbol interference.
The measurement noise is more than \num{10} times smaller than the minimum observed peak and thus there is no significant distortion.
Considering the injected volume per bit of \SI{14}{\micro\liter} and the stored volume of several \si{\milli\liter} in the syringe, several thousands of consecutive transmissions can be realized without refilling.
We also observe small variations in the amplitude of the peaks.
These can presumably be attributed to the expected variations of the volume injected by the peristaltic pump which mechanically compresses the particle injection tube.
On the other hand, the overall shape of the pulses is not affected by these variations.

\begin{figure}[!t]
    \centering
    \includegraphics{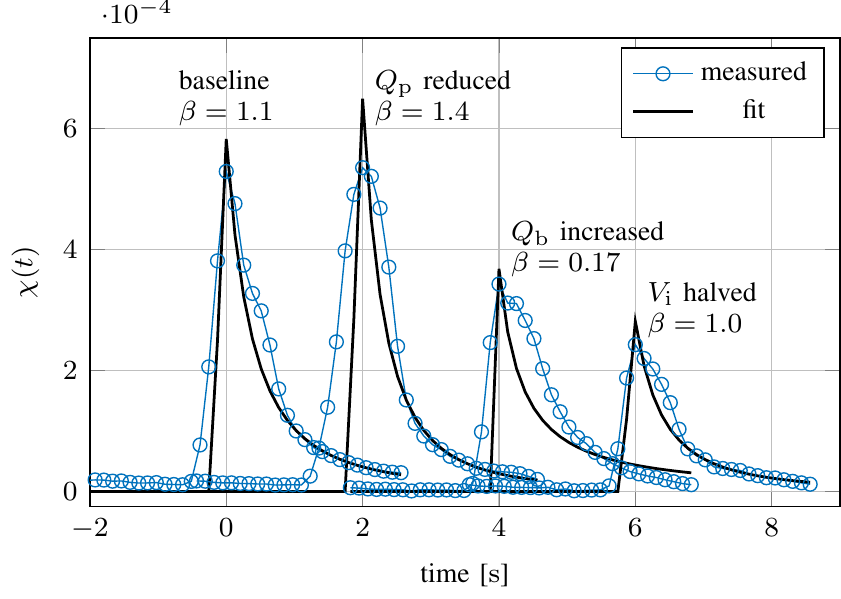}
    \caption{
        System response with fit according to \eqref{eq:system response}.
        For the baseline values, $\Qp=\SI{5.26}{\milli\liter/\minute}$, $\Qb=\SI{5}{\milli\liter/\minute}$, and $\Vi=\SI{14}{\micro\liter}$.
        Following baseline variations are considered.
        $\Qp$ reduced to \SI{1.4}{\milli\liter/\minute}, $\Qb$ increased to \SI{10}{\milli\liter/\minute}, and $\Vi$ halved to \SI{7}{\micro\liter}.
        Experimental data was averaged over four consecutive pulses.
    }
    \label{fig:base_ir}
    \vspace*{-0.25cm}
\end{figure}

In Fig.~\ref{fig:base_ir}, we show the measured susceptibility over time following a single injection.
We consider variations from the baseline parameters in Section~\ref{sec:system_description} regarding the injection flow rate $\Qp$, the background flow rate $\Qb$, and the volume of the injected particle suspension $\Vi$ as is also indicated in the figure.
Furthermore, we show the analytical solution $\chi(t)$ in \eqref{eq:system response}, where the shape parameter $\beta$ was numerically found by least-squares fitting to the measured data after time-shifting of the data such that the peaks of the measurement and of the analytical solution align.
For a better visualization, the curves are separated by \SI{2}{\second} each.
For all curves, we can observe a fast increase of the signal from $\chi=0$ to the peak value and a relatively slower decrease from the peak value back to $\chi=0$.
This behavior is also seen in the fitted analytical curve, which overall agrees reasonably well with the data.
Decreasing $\Qp$ to \SI{1.4}{\milli\liter/\minute} approximately preserves the amplitude of the pulse and also the decay after the peak with respect to the baseline value.
Correspondingly, the value $\beta=1.1$ of the fitted curve is similar to the baseline value $\beta=1.4$.
For times smaller than the peak time, the pulse is broader compared to the baseline value which is not captured by the analytical curve.
This can be explained by our modeling assumption of an axially concentrated release from the cross section which does not strictly hold.
Changing $\Qb$ to a larger value of \SI{10}{\milli\liter/\minute}, leads to a slower decay compared to the baseline scheme.
In this case, also the fit parameter is decreased from $\beta=1.1$ to $\beta=0.17$ which might be explained as follows.
When the background flow becomes stronger, during the injection duration, particles are transported away faster and might not reach the center of the tube where the velocity would be faster compared to when the background flow is slower.
Hence, more particles move with slower velocity and thus the measured signal decays slower.
When halving $\Vi$ to \SI{7}{\micro\liter}, the received signal is also approximately halved.
In this case, the initial release pattern is not yet heavily influenced by the reduction of $\Vi$ and parameter $\beta=1.0$ has a similar value as the baseline $\beta=1.1$.

\section{Conclusion}
\label{sec:conclusion}
We presented a new testbed for the investigation of molecular communications using the passive detection of engineered magnetic nanoparticles.
The proposed system enabled reliable communication, and the measurement of the magnetic susceptibility as a quantity proportional to the particle concentration allowed for a simple mathematical model based on laminar flow-driven particle transport.
Potential applications of magnetic nanoparticle based molecular communication include reporting sensing results and carrying control information in industrial, microfluidic or biomedical settings, especially at locations, where other forms of communication could not be employed.
The testbed could potentially be expanded by implementing a network of ducts, changing the carrier liquid, or scaling of its size.
Moreover, particles could be additionally tagged with other chemicals.
This would, for example, allow for distinguishing releases from different locations or for detecting which kind of event caused a release.



\bibliographystyle{IEEEtran}
\bibliography{spawc_bibliography}
\end{document}